\documentclass[reprint,aip,jap,amsmath,amssymb]{revtex4-1}
\usepackage{amsmath}
\usepackage{amsfonts}
\usepackage{amssymb}
\usepackage{graphics}
\usepackage{float}
\usepackage{graphicx}
\usepackage{epstopdf}
\usepackage[compact]{titlesec}
\usepackage{natbib}
\usepackage{threeparttable}
\usepackage[titletoc,title]{appendix}
\usepackage{lipsum}
\usepackage{varwidth}
\usepackage{tabularx}
\usepackage{siunitx}
\usepackage{gensymb}
\graphicspath{ {figures/Final_figures/} }

\newcommand{\ie}{\textit{i}.\textit{e}., }

\begin{document}
\title{The estimation of impact ionization coefficients for $\beta $-Ga$_{2}$O$_{3}$}
\author{Giftsondass Irudayadass}
\affiliation{Dept. of Electrical and Computer Engineering, University of Illinois, Chicago, IL 60607.}
\author{Junxia Shi}
\affiliation{Dept. of Electrical and Computer Engineering, University of Illinois, Chicago, IL 60607.}

\begin{abstract}
{Impact ionization coefficients of anisotropic monoclinic $\beta $-Ga$_{2}$O$_{3}$ are estimated along four crystallographic directions and the plot for the $\left[ 010 \right]$ direction is shown. The approximation models were fitted to Baraff's universal plot for ionization rate in semiconductors and the values were obtained for $\beta $-Ga$_{2}$O$_{3}$. The phonon mean free path of $\beta $-Ga$_{2}$O$_{3}$ was estimated to be 5.2604 \si{\angstrom} using Gray medium approximation. The phonon group velocity takes the value of longitudinal acoustic phonons. The ionization rate has a maximum value of 3.98\,$\times$\,10$^{6}$\,$cm^{-1}$ along the $\left[ 010 \right]$ direction over the applied electric field range (1.43-4)\,$\times10^{7}\,V\,\cdot\,cm^{-1}$. Contrary to expectations, the phonon mean free path along direction $\left[ \bar{2}01 \right]$ is the lowest since it has a lower thermal conductivity to phonon group velocity ratio. The plots were compared with GaN and 4H-SiC, which shows that as bandgap increases the field required for ionization increases. The critical electric field was estimated to be 0.921\,$\times10^{8}\,V\cdot\,cm^{-1}$ along $\left[ 010 \right]$ direction.}
\end{abstract}
\maketitle

\vspace{0.3cm}
\section{Introduction}
\vspace{0.3cm}
{

The semiconductor industry dominated by Si-based technology has also explored various materials and compounds for specific-targeted applications. A wide variety of device designs~\cite{doering2007handbook,o2007handbook} have been studied for applications as diverse as high-power switching~\cite{shenai1989optimum} to opto-electronic control~\cite{denbaars1997gallium, sengupta2011multiscale} of physical processes. However, there is still a need for improved material sets, for instance, silicon carbide (SiC) and wurtzite gallium nitride (GaN) are now well-established as foundational blocks for power electronic components. These compound semiconductors possess desired power electronics specific material properties, for example, a large band gap $ \left( E_{g}\right) $ and high breakdown field $ \left( E_{br}\right) $. A substantial body of work focussed on SiC and GaN power devices is available, however, there still lies areas where improved performance is desired. Recently, monoclinic Ga$_{2}$O$_{3} $ by virtue of its extraordinary material properties, an impressive Baliga's figure of merit $ \left(BFOM\right) $ (larger than those of Si, SiC and GaN) has renewed interest in applications centred around power switching, RF amplifiers, and in general signal processing under high electric field operational conditions. Its manifold uniqueness therefore warrants more accurate study of material physics, related device technologies, and an overall better understanding of its core set of physical properties. The intrinsic set of advantages notwithstanding and coupled to previous acquaintance with this material when they were bulk melt-grown as substrates, their commercial viability remains poor vis-\`a-vis silicon and their derivatives currently in use. A particularly attractive way of probing a  material deeper lies in the use of simulation tools that offset the cost of expensive experimental efforts. TCAD Simulation tools~\cite{fonseca2013efficient} help in reducing the cost while improving the quality of experimental research by highlighting device structures with desired output characteristics, but the majority of the tools and methods are oriented towards silicon based devices. While methods, models and coefficients, semi-empirical band-structure calculation techniques~\cite{klimeck2000si, sengupta2016numerical} are available for other materials like SiC, GaN and InP, little to no data are available for Ga$_{2}$O$_{3}$. In this work we will determine such coefficients and desirable models to be used for $\beta $-Ga$_{2}$O$_{3}$, particularly the coefficients for impact ionization modelling since the major focus on $\beta $-Ga$_{2}$O$_{3}$ is for power applications.

\begin{figure}[t]
\centering
\includegraphics[scale=1.9]{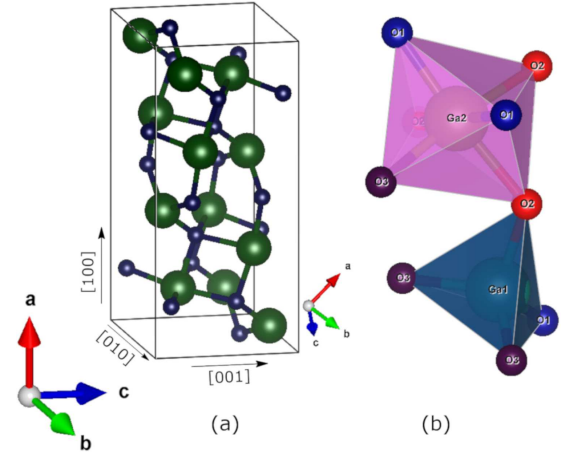}
\caption{\label{fig:upcell} The figure on the left panel, $\left(a\right)$, represents the unit cell of $\beta $-Ga$_{2}$O$_{3}$ which consists of two inequivalent Ga sites (large spheres), three inequivalent O-sites (small spheres) and contains 20 atoms (4 Ga$_{2}$O$_{3}$). The lattice parameters (length, breadth and width of the unit cell along the crystallographic axis, and the corresponding angles) are given as, a = 12.214\si{\angstrom}, b = 3.037\si{\angstrom}, c = 5.7981\si{\angstrom}, and $\beta$ = 103.83$\degree$ (between a and c)~\citep{geller1960crystal,aahman1996reinvestigation}. The figure on the right panel, $\left(b\right)$, shows the tetrahedral (Ga1) and octahedral (Ga2) geometry displayed by gallium atoms.}
\vspace{-0.5cm}
\end{figure}

Gallium Oxide single crystals exhibit polymorphism~\cite{roy1952polymorphism} with five confirmed polytypes $\left(\alpha,\beta,\gamma,\delta,\epsilon\right)$, of which the anisotropic monoclinic $\beta $-Ga$_{2}$O$_{3}$ is the most stable polymorph, thermodynamically. The unit cell of $\beta $-Ga$_{2}$O$_{3}$, shown in Fig~\ref{fig:upcell}$\left(a\right)$, consists of two inequivalent Ga sites (large spheres), three inequivalent O-sites (small spheres) and contains a total of 30 atoms and the primitive cell comprises 10 atoms. The tetrahedral and octahedral geometry of Ga1 and Ga2 sites are illustrated in Fig~\ref{fig:upcell}$\left(b\right)$. A detailed geometry as given in Ref.~\onlinecite{aahman1996reinvestigation,geller1960crystal} is illustrated in Appendix~\ref{geometry}. The lattice parameters corresponding to the crystallographic axis of $\beta $-Ga$_{2}$O$_{3}$ unit cell are given as, a = 12.214 \si{\angstrom}, b = 3.037 \si{\angstrom}, c = 5.7981\si{\angstrom}, and $\beta$ = 103.83$\degree$ (between a and c)~\citep{aahman1996reinvestigation}, which implies that this crystal belongs to the C12/m1 space group for which the parallelpiped reciprocal lattice vector unit cell Brillouin zone schematic is shown in Fig~\ref{fig:brillouin},~\cite{bradley2010mathematical,aroyo2006bilbao1,aroyo2006bilbao,aroyo2011crystallography,aroyo2014brillouin} and a detailed Brillouin zone computed using the primitive cell was suggested by H. Peelaers and C. G. Van de Walle~\cite{peelaers2015brillouin}. 

The dielectric constant of $\beta $-Ga$_{2}$O$_{3}$ was set to 10.2, as given in Refs.~\onlinecite{passlack1994dielectric,hoeneisen1971permittivity}, and the experimentally obtained band gap is ultra wide~\cite{janowitz2011experimental,tippins1965optical,matsumoto1974absorption,orita2000deep} at 4.85 $ \pm $ 0.1 eV and has an experimental electrical breakdown field of 3.6 \si{MV.cm^{-1}} ~\cite{passlack1995ga2o3} though the theoretical limit was estimated to be 8 \si{MV.cm^{-1}},~\cite{hudgins2003assessment} which encourages researchers towards the development of Ga$_{2}$O$_{3}$ devices, specifically for its applications in power electronics. Experimental estimates measure the effective electron mass $m_{e}$ to be around 0.3$m_{0}$~\cite{binet1994relation} and combining this with the first principle calculations a good compromise for $m_{e}$ will be 0.28$m_{0}$,~\cite{varley2010oxygen,yamaguchi2004first,he2006first} from which the conduction band density of states is be calculated as N$_{c}$= 3.7$\times$10$^{18}$ $cm^{3}$.~\cite{irmscher2011electrical} The $\beta $ polymorph of Ga$_{2}$O$_{3}$ has been studied extensively over other polytypes owing to its thermal and chemical stability and ease of substrate preparation from melt based growth techniques, and the thermal conductivity measured to be 27$ \pm $ 2.0, 14.7$ \pm $1.5, 13.3$ \pm $1.0 and 10.9$ \pm $ 1.0 W/m-K for crystallographic directions $ \left[010\right] $, $ \left[001\right] $, $ \left[\bar{2}01\right] $ and $ \left[100\right] $ respectively,~\cite{guo2015anisotropic} is different along different crystallographic axis, due to the crystal's anisotropic nature.

\indent Due to the lack of fundamental research, the estimation of impact ionization coefficients, namely ionization rate ($\alpha$), critical electric field ($E^{crit}$), applied electric field ($\mathbf{E}$) and ionization rate constant ($\alpha_{0}$), depends on the determination of phonon mean free path and fortunately enough data is available to estimate this value. Phonon mean free path was calculated to be 5.2604, 3.0675, 2.99 and 2.736 \si{\angstrom}, for crystallographic directions $ \left[010\right] $, $ \left[100\right] $, $ \left[001\right] $ and $ \left[\bar{2}01\right] $, respectively. The impact ionization rates were calculated using the standard impact ionization models $ \left(IIM\right) $ provided by Crowell-Sze, Sutherland and Thornber and these values were used in Selberherr IIM. The model suggested by Sutherland in Ref.~\onlinecite{sutherland1980improved} provides the best fit to Universal Baraff's Curve for ionization rates, and has the values 3.98 $ \times $  10$^{6}$, 7.626 $ \times $  10$^{6}$, 8.965 $ \times $  10$^{6}$ and 9.485 $ \times $  10$^{6}$ cm $^{-1}$ for crystallographic directions $ \left[010\right] $, $ \left[100\right] $, $ \left[001\right] $ and $ \left[\bar{2}01\right] $.} 


\begin{figure}[t]
\includegraphics[scale=0.35]{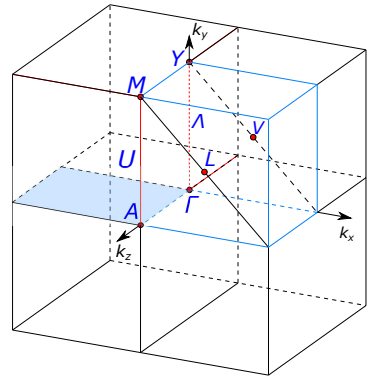}
\caption{\label{fig:brillouin} The k-vectors of the base-centered monoclinic $\beta $-Ga$_{2}$O$_{3}$ belonging to the arithmetic crystal class 12/$m$1$C$ (2/$mC$), space group $C$12/$m$1 (No. 12). The sketch is the axial ratio independent parallelepiped representation of the primitive unit cell, shown in Fig.~\ref{fig:upcell}.}
\vspace{-0.5cm}
\end{figure}

\vspace{0.3cm}
\section{Methods and Models} \label{Models}
\vspace{0.3cm}
Several impact ionization models can be used for the estimation of ionization rate, which consecutively aids in determining the breakdown voltage. Selberherr's model~\cite{siegfried1984analysis}, which is a modification of Chynoweth's law, takes the following expression, 
\begin{equation}
\alpha = \alpha_{0} \cdot \exp\left( -\left( \dfrac{E^{crit}}{\mathbf{E}}\right)^{\beta} \right), \label{eq:selb}
\end{equation} 
where, $\alpha$ is the ionization rate, $\alpha_{0}$ the impact ionization rate constant, $E^{crit}$ is the critical electric field, $\mathbf{E}$ is the applied electric field strength at a specific position in the direction of current flow, and $\beta$, a fitting constant, is in the range 1 to 2. The ionization coefficients for holes take similar values to that of electrons, for the purpose of simulations. To estimate the coefficients for Selberherr's model, the numerical approximation methods provided by Crowell-Sze, Sutherland and Thornber are fitted to Baraff's universal plot~\cite{baraff1962distribution} for ionization rate in semiconductors. Baraff plot follows the relation,
\begin{equation}
\alpha \cdot \lambda = f\left( \frac{E_{r}}{E_{i}},\frac{E_{i}}{q \cdot \lambda \cdot \mathbf{E}}\right), 
\end{equation}
where, $\alpha$ is the ionization rate, $\lambda$ the phonon mean free path, $E_{r}$ the optical phonon energy, $E_{i}$ the ionization energy. 
\\ The Baraff curve approximation proposed by Crowell and Sze~\cite{crowell1966temperature} is expressed as,
\begin{subequations} 
\begin{equation}
\alpha \cdot \lambda = \exp \left( R_{0}+ R_{1} \cdot x + R_{2} \cdot x^{2} \right),
\end{equation}
with:
\begin{align}
R_{0}&=-1.92+75.5 \cdot r-757 \cdot r^{2}, \\
R_{1}&=1.75 \cdot 10^{-2}-11.9 \cdot r+46 \cdot r^{2}, \\
R_{2}&=3.9 \cdot 10^{-4}-1.17 \cdot r+11.5 \cdot r^{2},
\end{align}
where,
\begin{align}
r &=\dfrac{E_{r}}{E_{i}}, \\ 
x &=\frac{E_{i}}{q \cdot \lambda \cdot \mathbf{E}} \cdot
\end{align}
\end{subequations}
This approximation is accurate over the range r $\in\left[ 0.01,0.06\right]$ and x $\in\left[ 5,16\right]$ within two percent maximum error. A more rigorous approximation was proposed by Sutherland~\cite{sutherland1980improved} given by, 

\begin{subequations}
\begin{equation}
\alpha \cdot \lambda =\exp ( R_{0}  + R_{1} \cdot x + R_{2}\cdot x^{2} + R_{3} \cdot x^{3} ),
\end{equation}
with:
\begin{equation}
R_{0}=-7.238 \cdot 10^{-2} -51.5 \cdot r + 239.6 \cdot r^{2} + 3357 \cdot r^{3},
\end{equation}
\begin{equation}
\begin{aligned}
R_{1}=-0.4844 + 12.45 \cdot r + 363 \cdot r^{2} - 5836 \cdot r^{3}, 
\end{aligned}
\end{equation}
\begin{equation}
R_{2}=2.982 \cdot 10^{-2} - 7.571 \cdot r -148.1 \cdot r^{2} + 1627 \cdot r^{3}, 
\end{equation}
\begin{equation}
R_{3}=-1.841 \cdot 10^{-5} - 0.1851 \cdot r + 10.41 \cdot r^{2} - 95.65 \cdot r^{3}.
\end{equation}
\end{subequations} 
For the range r $\in\left[ 0.01,0.07\right]$ and x $\in\left[ 3,14\right]$ this approximation is expected to fit Baraff's curve perfectly. The empirical expression proposed by Thornber~\cite{thornber1981applications} has been consistent with an elaborate momentum and energy scaling theory and is given by, 

\begin{subequations}
\begin{equation}
\alpha = \dfrac{\mathbf{E}}{E_{i}} \cdot \exp \left(- \dfrac{B_{j}}{\dfrac{k \cdot T \cdot B_{j}}{E_{i}} + \mathbf{E} + \dfrac{E^{2}}{B_{r}}} \right), 
\end{equation}
where,
\begin{align}
B_{j} &= \dfrac{E_{r}}{q \cdot \lambda}, \\
B_{r} &= \dfrac{E_{i}}{q \cdot \lambda} \cdot
\end{align}
\end{subequations}
$B_{j}$ is the threshold field at which the ionization energy is reached in one mean free path and $B_{r}$ is when phonon energy is reached in one mean free path. To determine the value of impact ionization rates from the above expressions, we need to determine the value of optical phonon mean free path (MFP), on which not much experimental or theoretical research has been done and no data is available for $\beta $-Ga$_{2}$O$_{3}$. The value of MFP can be determined from its relation to thermal conductivity and specific heat capacity provided as Gray's approximation,
\begin{equation}
\kappa \simeq \dfrac{1}{3} \cdot v_{ph} \cdot \lambda_{ph} \cdot c_{v}, \label{eq:kappa}
\end{equation}
where $\kappa$ is the thermal conductivity, $v_{ph}$ is the phonon group velocity (values for $v_{ph}$ is available in Ref~\onlinecite{guo2015anisotropic}), $\lambda_{ph}$ is the phonon mean free path, which needs to be determined, $c_{v}$ is the specific heat capacity which is calculated using the Debye model of specific heat represented as,
\begin{equation}
C_{v} = 3Nk\dfrac{3}{x_{D}^{3}}\int_{0}^{x_{D}} \dfrac{x^{4}e^{x}}{\left( e^{x}-1\right)^{2} } dx,
\end{equation}
where, \begin{scriptsize}$x_{D} = \dfrac{\theta_{D}}{T}$ \end{scriptsize} and $\theta_{D}$ is the Debye temperature, which can take the experimentally measured value of 738\,K~\cite{guo2015anisotropic} or the first principles estimate of 872\,K~\cite{he2006electronic} for $\beta $-Ga$_{2}$O$_{3}$, $T$ is the absolute temperature, $k$ is the Boltzmann constant and $N$ is the number of atoms, considered as Avogadro's number. We considered the experimentally measured value for Debye temperature.

\indent The models described above are pseudolocal in nature. Hence, the ionization rates calculated depend only on the electric field strength applied but not on the position where the carriers are generated. This suggests that the models can be used for any position within the space-charge region, which is not true. Such a description assumes an unphysical situation within the device for structures with sufficiently high multiplication. A critical multiplication ratio $\left( M_{c}\right)$, between the ionized and the total number of carriers at a any position in the structure, was suggested by Okuto and Crowell in Ref.~\onlinecite{okuto1974ionization} to address this pseudolocal issue in their non-localized concept description of avalanche ionization effect. M$_{c}$ is given by the following relation,

\begin{subequations}
\begin{equation}
M_{c}=\dfrac{2[1 + \alpha_{r}(\mathbf{E})D]}{\alpha_{a}(\mathbf{E})D[2+\alpha_{r}(\mathbf{E})D]}
\end{equation}
with:
\begin{equation}
\alpha_{a}(\mathbf{E})=\dfrac{\alpha_{r}(\mathbf{E})[1+\alpha_{r}(\mathbf{E})D]}{1+4\alpha_{r}(\mathbf{E})D+2\alpha_{r}^{2}(\mathbf{E})D^{2}},
\end{equation}
\begin{equation}
\alpha_{r}(\mathbf{E})=[X-D]^{-1},
\end{equation}
\begin{equation}
D=\dfrac{E_{i}-N_{r}E_{r}}{q \cdot \mathbf{E}},\\
\end{equation}
\begin{equation}
\begin{aligned}
X &=Dexp(((D/\lambda_{r})^{2}+[0.217(E_{i}/E_{r})^{1.14}]^{2})^{1/2} \\
  & -0.217(E_{i}/E_{r})),
\end{aligned}
\end{equation}
\end{subequations}

\indent where $N_{r}$ is the net number of optical phonons absorbed by the carrier and is assumed as zero, since only absorption of energy is considered at low temperatures. \textit{X}, is the average distance at which an ionizatoin scattering occurs, \textit{D}, is the dark-space distance, $\alpha_{r}(\mathbf{E})$ is the non-localized single-carrier ionization probability, $\alpha_{a}(\mathbf{E})$ is the ``apparent ionization coefficient" introduced in Ref~\onlinecite{okuto1974ionization}. The expression for $M_{c}$ has the following theoretical limits,
\begin{subequations}
\begin{equation}
M_{c} \to \dfrac{1}{\alpha_{a}(\mathbf{E})D} \quad when  \quad  \mathbf{E} \to 0,
\end{equation}

\begin{equation}
M_{c} \to 4 \quad when  \quad  \mathbf{E} \to \infty.
\end{equation}
\end{subequations}

\vspace{0.3cm}
\section{Results}
\vspace{0.3cm}

{In order to estimate the impact ionization coefficients we first determine the value for optical phonon mean free path ($\lambda_{ph}$) using Eq.~\ref{eq:kappa} where the values used for thermal conductivity, phonon group velocity and specific heat capacity are listed in Table~\ref{table1}. Owing to anisotropy in the crystal structure, the approximate values of $\lambda_{ph}$ are obtained for different crystallographic directions as catalogued in Table II. These estimates are justified when we compare the thermal conductivities of $\beta $-Ga$_{2}$O$_{3}$ and GaN given as 27 and 230 W/m-K~\cite{jezowski2003thermal}, respectively. The phonon group velocity and specific heat capacity of GaN is determined to be 6.9-8.2\,$\times$\,$10^{5}cm\cdot s^{-1}$,~\cite{truell2013ultrasonic} and 0.49 $J\cdot$ $g^{-1}$ $\cdot$\degree $C^{-1}$,~\cite{levinshtein2001properties} respectively. Corresponding values for $\beta $-Ga$_{2}$O$_{3}$ are 7.8\,$\times$\,$10^{5}cm\cdot s^{-1}$ and 0.56 $J\cdot$ $g^{-1}$ $\cdot$\degree$C^{-1}$ (for $\left[ 010 \right]$ direction)~\cite{galazka2014bulk}. The difference is $\sim$5\% for phonon group velocity and $\sim$13\% for specific heat capacity between the two compound semiconductors. Using Eq.~\eqref{eq:kappa} we see that the phonon mean free path is the only variable affecting thermal conductivity and has to be larger in GaN than $\beta $-Ga$_{2}$O$_{3}$. In fact $\lambda_{ph}$ of GaN is larger by an order of 10 from Ref~\onlinecite{danilchenko2006heat}. We should note that the $\lambda_{ph}$ thus obtained is for the acoustic branch. We consider a situation where no acoustic phonons are present. If the energy of the electron is below the optical phonon energy the mean free path will be infinite since no collision will take place unless the electron gains energy above the optical phonon energy. The presence of acoustic phonons gives a maximum value for $\lambda_{ph}$ beyond which it cannot increase, and this estimation will be a good approximation for the purpose of calculating the ionization coefficients. The value of $\lambda_{ph}$ is estimated to be 5.2604 \si{\angstrom} for $\beta $-Ga$_{2}$O$_{3}$ along the $ \left[010\right] $ direction, and $E_{r}$ takes the value 24.81 meV from Ref.~\onlinecite{kranert2016raman}.

\begin{figure}[t]
\begin{flushleft}
\includegraphics[scale=0.547]{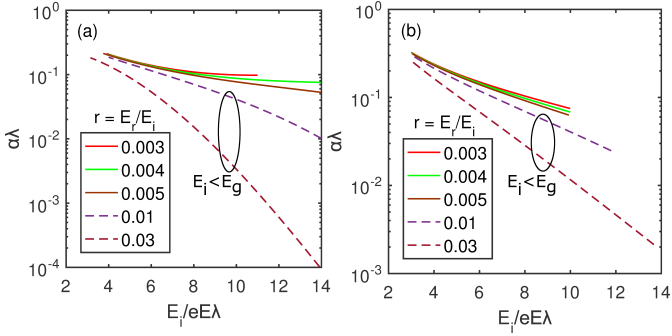}
\centering
\caption{\label{fig:baraff} The numerically obtained Baraff plots for ionization rates in $\beta $-Ga$_{2}$O$_{3}$ along the $ \left[010\right] $ direction using Sutherland model (figure on the left panel) and Thornber model (figure on the right panel). The plots were obtained for different values of ionization energy $ \left( E_{i}\right) $, including values of $E_{i}$ extended below $E_{g}$ ( plotted as dashed lines) to show that the obtained curves fit the universal Baraff plot for semiconductors, though physically it is impossible for ionization to take place. The range of the applied electric field to which the models fit is assumed to be the range  }
\end{flushleft}
\vspace{-0.5cm}
\end{figure}

\begin{table}[b]
\caption{Thermal Conductivity and Phonon group velocity of $\beta $-Ga$_{2}$O$_{3}$ for different crystallographic directions from literature ~\cite{guo2015anisotropic}.}
\centering
\label{table1}
\begin{tabular}{m{2.5cm} m{2cm} m{2cm} }
\noalign{\smallskip} \hline \hline \noalign{\smallskip}
Crystallographic direction & Thermal Conductivity $\left( W/m-K \right)$  & Phonon group velocity$\left(L_{A}\right)$ $\left(10^{5}cm\cdot s^{-1}\right)$ \\\hline
\vspace{0.1cm}
$[010]$  & 27$ \pm $ 2.0  & 7.8 \\
$[100]$  & 10.9$ \pm $1.5 & 5.4 \\
$[001]$  & 14.7$ \pm $1.0 & 7.1  \\
$[\bar{2}01]$ & 13.3$ \pm $1.0 & 6.6   \\

\noalign{\smallskip} \hline \noalign{\smallskip}
\end{tabular}
\end{table}

\begin{table}[b]
\caption{Impact ionization coefficients for different crystallographic directions of $\beta $-Ga$_{2}$O$_{3}$. Values of GaN (from Ref.~\onlinecite{danilchenko2006heat,jezowski2003thermal,levinshtein2001properties,ouguzman1997theory,truell2013ultrasonic}) and 4H-SiC (from Ref.~\onlinecite{choyke1969optical,berger1996semiconductor,harris1995properties,konstantinov1997ionization}) is given for comparison.}
\centering
\label{table2}
\begin{tabular}{m{1.4cm} m{1.6cm} m{1.8cm} m{1.6cm} m{1.5cm} m{2.5cm}}
\noalign{\smallskip} \hline \hline \noalign{\smallskip}
Crystallo graphic direction (Material) & Impact ionization rate ($\alpha$) [$\times10^{6}cm^{-1}$] & Critical Electric Field ($E^{crit}$) [$\times10^{8}Vcm^{-1}$] & Phonon Mean Free Path ($\lambda$)[\si{\angstrom}] & Applied Electric Field$\left( \mathbf{E} \right)$[ $\times10^{7}Vcm^{-1}$ ]\\\hline
\vspace{0.1cm}
$[010]$  & 3.98 & 0.921 & 5.260 & 1.43-4 \\
$[100]$  & 7.626 & 1.581 & 3.067 & 3.37-7.7\\
$[001]$  & 8.965 & 1.622 & 2.99 & 3.08-9.2   \\
$[\bar{2}01]$ & 9.485 & 1.755 & 2.763 & 3.5-9.7    \\

\noalign{\smallskip} \hline \noalign{\smallskip}
(GaN) & 0.25 & 0.34 & 580 & 0.12-0.53 \\
(4H-SiC) & 0.15 & 0.16 & 32.5 & 0.09-0.35\\

\noalign{\smallskip} \hline \noalign{\smallskip}
\end{tabular}
\end{table}

\begin{figure}[t]
\begin{flushleft}
\includegraphics[scale=0.68]{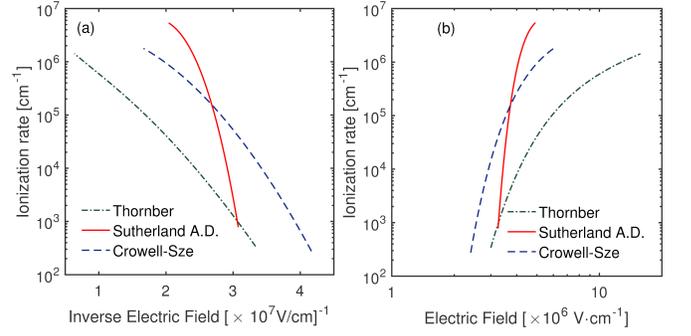}
\centering
\caption{\label{fig:E_Field} The theoretically calculated ionization rate curves $ \left( \alpha\right) $ for $\beta $-Ga$_{2}$O$_{3}$ at 300\,K along the $ \left[010\right] $ direction for all three methods. The rates calculated are for both electrons and holes. The figure on the left panel gives the ionization coefficient $\left( \alpha\right)$ as a function of inverse electric field, from where the ionization coefficient is derived and the figure on the right panel shows the ionization rate $\left( \alpha\right)$ as a function of electric field from which the applied electric field strength range is obtained. The previously determined ionization rate is valid within this range. }
\end{flushleft}
\vspace{-0.345cm}
\end{figure}

\indent Baraff adopts an approach which neither relies on diffusion approximation followed by Wolff~\cite{wolff1954theory} nor Shockley's~\cite{shockley1961problems} ``spike" distribution to describe electron transport but rather derives an integral equation for the collision density in order to estimate ionization rates, for all semiconductors. The curves from the mathematical models discussed in section~\ref{Models} must fit this universal plot provided by Baraff for different values of ionization energy ($E_{i}$) over the range of applied electric field strength $\left( \mathbf{E} \right)$. The values to which they fit for $\beta $-Ga$_{2}$O$_{3}$ are given in Table~\ref{table2}. The Baraff plots illustrated are for Sutherland model, Fig.\ref{fig:baraff}$\left( a \right)$, and Thornber model, Fig.\ref{fig:baraff}$\left( b \right)$, from which we can observe that Sutherland model fits more perfectly and for a larger range than Thornber model. Thus the values of ionization rate $\left( \alpha \right)$  and applied electric field strength $\left( \mathbf{E} \right)$ are calculated using Sutherland model. The solid lines are for values where the ionization energy is above the bandgap energy $(E_{i} > E_{g})$ and the dashed lines are for values of $E_{i}$ below $E_{g}$. The probability of ionization is less for values of $E_{i}$ below $E_{g}$, but are shown to clarify that the approximation models fit the Baraff curve. The range in which the values for ionization rate, $\alpha$, and applied electric field strength $\left( \mathbf{E} \right)$ are reliable for the material under consideration can also be extracted from Baraff curve. The plot for the model provided by Crowell-Sze is not shown. The figures are plotted only for $\left[ 010 \right]$, since the thermal conductivity is maximum along this direction. The estimates of the coefficients $\alpha$ and $\left( \mathbf{E} \right)$ are extracted from Fig.~\ref{fig:E_Field}$\left( a \right)$ (ionization coefficient $vs$ inverse electric field) and Fig.~\ref{fig:E_Field}$\left( b \right)$ (log-log plot) respectively. The ionization rate constant, $\alpha_{0}$, can be calculated from these values using Eq.~\eqref{eq:selb}. The values of $\lambda_{ph}$, $E_{r}$ and $E_{i}$ used to plot Fig.~\ref{fig:E_Field} are 5.2604\si{\angstrom}, 24.81 meV and 7.275 eV (\ie 1.5 $\times$ $E_{g}$), respectively. There are no experimental verification available for ionization rates of $\beta $-Ga$_{2}$O$_{3}$.

\begin{figure}[t]
\includegraphics[scale=0.68]{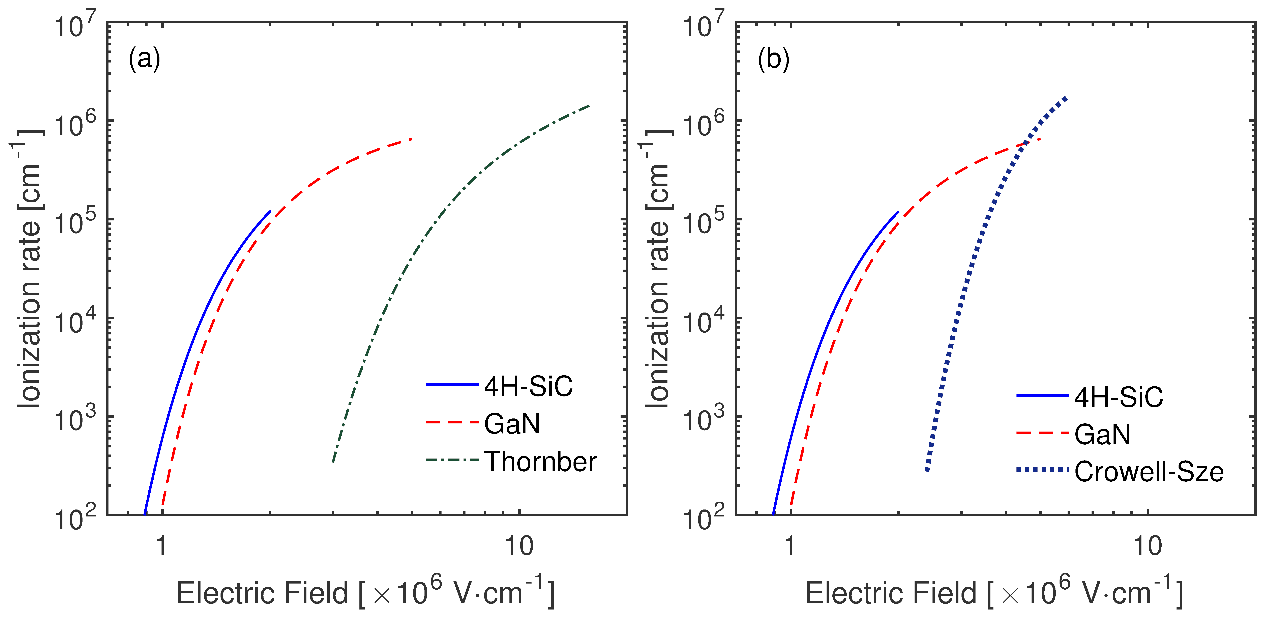}
\caption{\label{fig:Comparison} Ionization rate curves for 4H-SiC, GaN and $\beta $-Ga$_{2}$O$_{3}$. Comparison of the curves suggests that the electric field strenght predominantly depends on the bandgap and the predicted ionization rate would be higher in $\beta $-Ga$_{2}$O$_{3}$. The curves for GaN were traced using parameters from references~\onlinecite{danilchenko2006heat,jezowski2003thermal,levinshtein2001properties,ouguzman1997theory,truell2013ultrasonic} and for 4H-SiC from~\onlinecite{choyke1969optical,berger1996semiconductor,harris1995properties,konstantinov1997ionization}. The ionization rate plot of GaN was extrapolated for clarity. }
\vspace{-0.5cm}
\end{figure}

 The ionization rate constant can be estimated from its relationship to ionization rate, critical field and applied field as shown in Eq.~\eqref{eq:selb}. The range of values thus obtained are for a range of the electric field estimated from Baraff plot. The exact value of ionization rate constant, $\alpha_{0}$, can only be determined through experimentation, or at the least by knowing the breakdown voltage and related electric field applied to the device under consideration. We predict the ionization rate constant to be (\,2.51\,$\times$\,10$^9$\,-\,3.9\,$\times$\,10$^7$)\,cm$^{-1}$ over the applied electric field range of (\,1.43\,$\times$\,10$^7$\,-\,4\,$\times$\,10$^7$)\,V\,$\cdot$\,cm$^{-1}$. The predicted values for ionization coefficients are listed in Table.~\ref{table2} and given the large energy band gap of $\beta $-Ga$_{2}$O$_{3}$, the value estimated for applied electric field strength can be justified when juxtaposed with other wide-bandgap semiconductors.

\begin{figure}[t]
\includegraphics[scale=0.7]{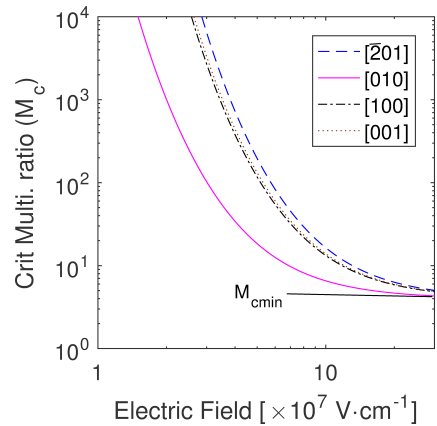}
\caption{\label{fig:CritMul} The critical multiplication ratio (M$_c$) as a function of electric field, for both electrons and holes, along the four crystallographic directions at 300\,K. M$_{cmin}$ denotes the theoretical minimum value for M$_{c}$ at infinite field. M$_{c}$ is the ratio of the total number of carriers (electrons + holes) to the number of electrons (or holes). If the ratio of the total number of carriers to the number of ionized carriers is higher than M$_{c}$ the approximations do not hold.}
\vspace{-0.55cm}
\end{figure}

\indent A comparison of $\alpha (vs) \mathbf{E}$ plots between major wide-bandgap semiconductor materials considered for power electronics can be derived from Fig.~\ref{fig:Comparison}. The curves were traced using the variables described in section~\ref{Models} with parameters from Ref.~\onlinecite{danilchenko2006heat,jezowski2003thermal,levinshtein2001properties,ouguzman1997theory,truell2013ultrasonic} for GaN, Ref.~\onlinecite{choyke1969optical,berger1996semiconductor,harris1995properties,konstantinov1997ionization} for 4H-SiC. The applied field strength range at which ionization occurs is larger in $\beta $-Ga$_{2}$O$_{3}$ than GaN or 4H-SiC and a similar comparison can be made for materials germanium, silicon, gallium arsenide and gallium phosphide from Ref.~\onlinecite{okuto1972energy}. The increase in applied field is attributed to the bandgap of the material since the field strength required for ionization increases as the bandgap increases. The ionization rate depends on the values of phonon mean free path $\lambda_{ph}$ and optical phonon energy $E_{r}$. The phonon mean free path determines, the average distance a carrier has to travel to acquire enough energy for ionization and $E_{r}$ influences the ratio of ``cross section" ($r$) for ionization. Comparing $\lambda_{ph}$ and $E_{r}$ of gallium oxide to corresponding values of GaN and 4H-SiC predicts the ionization rate to be higher in $\beta $-Ga$_{2}$O$_{3}$, as estimated. The critical multiplication ratio $M_{c}$ discussed in section~\ref{Models} aids in understanding the limitations of the ionization models and the numerically calculated $M_{c}$ is shown in Fig.~\ref{fig:CritMul}. If the ratio of the total number carriers to the ionized carrier is higher than $M_{c}$ then the models do not hold and if it is lower the boundary conditions are important. We can see that $M_{c}$ has a large value at low field and saturates at high field as predicted by the theoretical limits and also notice that the field strength at which the multiplication holds is high and compliments the field strength obtained from Baraff plots.
}

\vspace{0.5cm}
\section{Conclusion}
\vspace{0.3cm}
{ We have estimated the ionization coefficients of $\beta $-Ga$_{2}$O$_{3}$ using approximation models provided by Crowell-Sze, Thornber and Sutherland, for Baraff's universal plot of ionization rates in semiconductors. The phonon mean free path was estimated using Gray approximation of thermal conductivity and debye's model was used to determine specific heat. The ionization rate curves thus determined were compared with other major power semiconductors (GaN and 4H-SiC), and it was found that as the bandgap increases the field strength required for ionization also increases, regardless of phonon energy. Values for critical multiplication ratio of the carriers was determined to address the pseudolocal nature of the approximations. A plot for the same is illustrated and above these values the models do not hold.

The ionization rate constant can be determined by measuring the breakdown voltage and the electric field at this breakdown. Using the measured applied field, in the expression for ionization rate, we can deduce the ionization rate constant. Alternatively, existing breakdown values of gallium oxide devices can be used, provided we know the crystal orientation of the channel of the device during breakdown. Then this device can be simulated for the suggested range of values of ionization rate until the measured breakdown voltage is obtained. Finally, the ionization rate constant can be calculated using its relation to ionization rate and applied electric field.

The approximation models considered in this work, although developed based on silicon and gallium arsenide, holds true for a wide bandgap semiconductor, such as gallium oxide, since the carrier density approach by Baraff is universal for all semiconductors. The dependence of ionization rate to the change in carrier concentration was not considered. An experimental analysis measuring I$_{d}$-V$_{d}$ characteristics of $\beta $-Ga$_{2}$O$_{3}$ MOSFET, such as in Ref.~\onlinecite{higashiwaki2013depletion} and recording information on breakdown voltage and applied electric field, will assist in verifying the estimated ionization coefficient values.}


\vspace{0.5cm} 
\begin{appendices}
\vspace{0.5cm}
\section{Geometry of $\beta $-Ga$_{2}$O$_{3}$} \label{geometry}
\begin{figure}[htb!]
\begin{flushleft}
\includegraphics[scale=0.14]{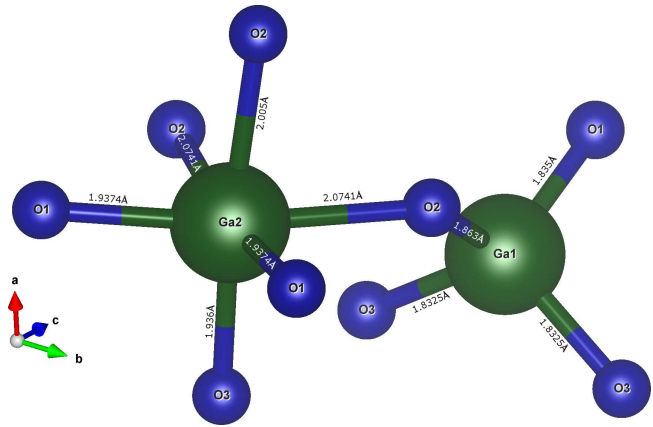}
\centering
\caption{\label{fig:bondlength} The both lengths between the atoms are illustrated in this figure. Ga1 and Ga2 are the tetrahedral and octahedral sites, respectively. }
\end{flushleft}
\vspace{-0.5cm}
\end{figure}

\begin{figure}[htb!]
\begin{flushleft}
\includegraphics[scale=2]{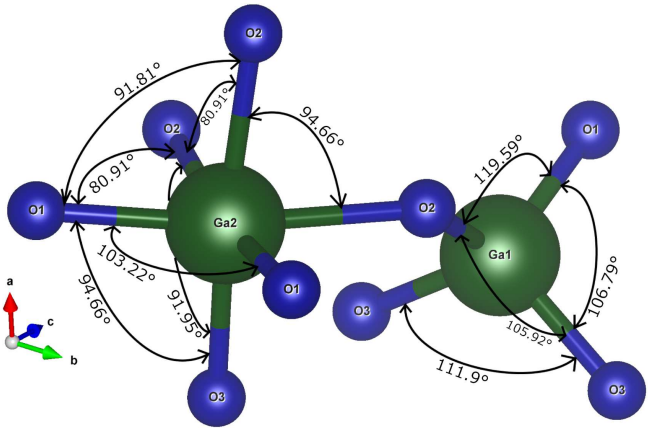}
\centering
\caption{\label{fig:bond_angle} The angles between the bonds are illustrated in this figure. Ga1 and Ga2 are the tetrahedral and octahedral sites, respectively. }
\end{flushleft}
\vspace{-0.5cm}
\end{figure}

\vspace{0.5cm}
The bond lengths and angles for the two Ga sites are given in Table~\ref{bondlength}\,and\,\ref{angles} and illustrated in Fig.~\ref{fig:bondlength}\,and\,\ref{fig:bond_angle}.

\begin{table}[t!]
\caption{Fractional atomic coordinates and isotropic displacement parameters of atoms in $\beta $-Ga$_{2}$O$_{3}$ from Ref.~\onlinecite{aahman1996reinvestigation}.}
\centering
\label{coordinates}
\begin{tabular}{m{2cm} m{1.5cm} m{0.5cm} m{1.5cm} m{1.5cm}}
\noalign{\smallskip} \hline \hline \noalign{\smallskip}
Element & x & y & z & $U_{eq}$ \\\hline
\vspace{0.1cm}
Ga1  & 0.09050(2) & 0 & 0.7946(5) & 0.0038(1) \\
Ga2  & 0.15866(2) & $1/2$ & 0.31402(5) & 0.0040(1) \\
O1   & 0.1645(2) & 0 & 0.1098(3) & 0.0060 (4)  \\
O2   & 0.1733(2) & 0 & 0.5632(4) & 0.0056 (4)   \\
O3   & $-0.0041(2)$ & $1/2$ & 0.2566(3) & 0.0042(4)       \\
\noalign{\smallskip} \hline \noalign{\smallskip}
\end{tabular}
\end{table}

\begin{table}[htb!]
\caption{The bond lengths and angles between atoms in $\beta $-Ga$_{2}$O$_{3}$ from Ref.~\onlinecite{aahman1996reinvestigation,geller1960crystal}.}
\centering
\label{bondlength}
\begin{tabular}{m{2.5cm} m{2.5cm} }
\noalign{\smallskip} \hline \hline \noalign{\smallskip}
Bonds & Bond length (\si{\angstrom}) \\\hline
\vspace{0.1cm}
Gal--O1$^{i}$   & 1.835(2) \\
Ga1--O2         & 1.863(2) \\
Gal--O3$^{ii}$  & 1.833(1)  \\
Ga2--O1         & 1.937(1)  \\
Ga2--O2         & 2.074(1)  \\
Ga2--O2$^{iii}$ & 2.005(2) \\
Ga2--O2         & 1.935(2)  \\
\noalign{\smallskip} \hline \noalign{\smallskip}
\end{tabular}
\end{table}


The equivalent isotropic displacement parameter, $U_{eq}$, is given by,
\begin{equation}
U_{eq}=(\dfrac{1}{3}) \Sigma_{i}\Sigma_{j} U_{ij} a_{i}^{*} a_{j}^{*} \bm{a_{i}}\,\cdot\,\bm{a_{i}}
\end{equation}

\begin{table}[htb!]
\caption{The bond lengths and angles between atoms in $\beta $-Ga$_{2}$O$_{3}$ from Ref.~\onlinecite{aahman1996reinvestigation,geller1960crystal}.}
\centering
\label{angles}
\begin{tabular}{m{2.5cm} m{2cm} }
\noalign{\smallskip} \hline \hline \noalign{\smallskip}
Bonds & Angles($\degree$) \\\hline
\vspace{0.1cm}
O1$^{i}$--Gal--O2         & 119.59(9)  \\
O1$^{i}$--Gal--O3$^{ii}$  & 106.79(7) \\
O2$^{i}$--Ga1--O3$^{ii}$  & 105.92(7)  \\
O3$^{ii}$--Gal--O3$^{iv}$ & 111.9(1)  \\
O1$^{i}$--Ga2--O1$^{v}$   & 103.22(9)  \\
O1--Ga2--O2               & 80.91(6) \\
O1$^{i}$--Ga--O2$^{iii}$  & 91.87(7)  \\
O1--Ga2--O3               & 94.66(7)  \\
O2--Ga2--O2$^{v}$         & 94.14(7)  \\
O2--Ga2--O2$^{iii}$       & 80.91(6)  \\
O2--Ga2--O3               & 91.95(7)  \\
\noalign{\smallskip} \hline \noalign{\smallskip}
\end{tabular}
\end{table}

The symmetric codes used in Table~\ref{bondlength}and~\ref{angles} are as follows,

(i) x,y,1+z; (ii) -x,-y,1-z; (iii) $\dfrac{1}{2}$-x,$\dfrac{1}{2}$-y,1-z;
(iv) -x,1-y,1-z; (v) x,1+y,z.


\end{appendices}


\end{document}